\documentclass[preprint,amsmath,showpacs]{revtex4}
\def\DESepsf(#1 width #2){\epsfxsize=#2 \epsfbox{#1}}
\usepackage{graphicx}
\usepackage{color}
\usepackage{amsmath}
\def\bmatrix{\left[\begin{array}}
\def\ematrix{\end{array}\right]}

\begin{document}

%

\let\a=\alpha      \let\b=\beta       \let\c=\chi        \let\d=\delta
\let\e=\varepsilon \let\f=\varphi     \let\g=\gamma      \let\h=\eta
\let\k=\kappa      \let\l=\lambda     \let\m=\mu
\let\o=\omega      \let\r=\varrho     \let\s=\sigma
\let\t=\tau        \let\th=\vartheta  \let\y=\upsilon    \let\x=\xi
\let\z=\zeta       \let\io=\iota      \let\vp=\varpi     \let\ro=\rho
\let\ph=\phi       \let\ep=\epsilon   \let\te=\theta
\let\n=\nu
\let\D=\Delta   \let\F=\Phi    \let\G=\Gamma  \let\L=\Lambda
\let\O=\Omega   \let\P=\Pi     \let\Ps=\Psi   \let\Si=\Sigma
\let\Th=\Theta  \let\X=\Xi     \let\Y=\Upsilon

%

%

\def\cA{{\cal A}}                \def\cB{{\cal B}}
\def\cC{{\cal C}}                \def\cD{{\cal D}}
\def\cE{{\cal E}}                \def\cF{{\cal F}}
\def\cG{{\cal G}}                \def\cH{{\cal H}}
\def\cI{{\cal I}}                \def\cJ{{\cal J}}
\def\cK{{\cal K}}                \def\cL{{\cal L}}
\def\cM{{\cal M}}                \def\cN{{\cal N}}
\def\cO{{\cal O}}                \def\cP{{\cal P}}
\def\cQ{{\cal Q}}                \def\cR{{\cal R}}
\def\cS{{\cal S}}                \def\cT{{\cal T}}
\def\cU{{\cal U}}                \def\cV{{\cal V}}
\def\cW{{\cal W}}                \def\cX{{\cal X}}
\def\cY{{\cal Y}}                \def\cZ{{\cal Z}}
%

\newcommand{\Ns}{N\hspace{-4.7mm}\not\hspace{2.7mm}}
\newcommand{\qs}{q\hspace{-3.7mm}\not\hspace{3.4mm}}
\newcommand{\ps}{p\hspace{-3.3mm}\not\hspace{1.2mm}}
\newcommand{\ks}{k\hspace{-3.3mm}\not\hspace{1.2mm}}
\newcommand{\des}{\partial\hspace{-4.mm}\not\hspace{2.5mm}}
\newcommand{\desco}{D\hspace{-4mm}\not\hspace{2mm}}
\renewcommand{\figurename}{Fig.}


%
\title{\boldmath
Sensitivities of IceCube DeepCore Detector to Signatures of Low-Mass Dark Matter in the Galactic Halo  }
\vfill
\author{Fei-Fan Lee$^{1}$}
\author{Guey-Lin Lin$^{1}$}
\author{Yue-Lin Sming Tsai$^{2,3}$}
\affiliation{
 $^1$Institute of Physics,
 National Chiao-Tung University,
 Hsinchu 30010, Taiwan
}
\affiliation{
$^2$National Centre for Nuclear Research,
Ho$\dot{z}$a 69, 00-681 Warsaw, Poland}
\affiliation{
$^3$Kavli Institute for Theoretical Physics, CAS, Beijing 100190, China}
\date{\today}
%
%
%
\begin{abstract}

We discuss the event rate in DeepCore array due to neutrino flux produced by
annihilations and decays of galactic dark matter. This event rate is calculated with a 10 GeV threshold energy, which is smaller than
 the threshold energy taken in previous works.
Taking into account the background event rate due to the atmospheric neutrino flux, we  
evaluate the sensitivity of DeepCore array for probing dark matter annihilation cross section and decay time.  
The sensitivity studies include the annihilation modes $\chi\chi\to b\bar{b}, \
\tau^+ \tau^-$, \ $\mu^+\mu^-$, and $\nu\bar{\nu}$, and decay modes $\chi\to b\bar{b}, \ \tau^+ \tau^-$, \ $\mu^+\mu^-$, and  
$\nu\bar{\nu}$. We compare our results with corresponding constraints derived from observations of WMAP, ACT and Fermi-LAT. 

\end{abstract}
%
\maketitle

\pagestyle{plain}

\section{Introduction}

Many astrophysical observations indicate the existence of dark matter (DM). A good example of such observations  is the measurement of  rotation curves for stars and gas in spiral galaxies.   
On the other hand, the nature of DM remains to be unveiled.  In this regard, many DM candidates have been proposed with the proposal of weak interacting massive particles (WIMPs)  \cite{Jungman:1996,Bertone:2005} the most popular among all candidates.  The detections of DM shall test the idea of WIMPs. The detections of DM can be categorized into direct and indirect approaches. 
The former approach proceeds by observing the nucleus recoil as DM interacts with the target nuclei in the detector. The latter approach relies on detecting 
final state particles resulting from DM annihilations or decays.   In this article, we focus on the indirect detection of DM through observing neutrinos produced by 
DM annihilations or decays in the galactic halo.  

The search for neutrinos coming from DM annihilations in galactic halo has been performed by IceCube \cite{Achterberg:2006}. Data obtained from IceCube 22-string configuration set the 90\% C.L. upper limit on DM annihilation cross section  $\langle\sigma\upsilon\rangle\sim 10^{-22}$cm$^3$s$^{-1}$  for $\chi\chi\to \nu\bar{\nu}$ channel  at $m_{\chi}=1$ TeV~\cite{Abbasi:2011eq}, while the preliminary result of IceCube 40-string galactic-center analysis improves the above limit to $10^{-23}$cm$^3$s$^{-1}$~\cite{IceCube:2011ae}. Based upon IceCube 40-string data selection for searching diffuse flux of astrophysical muon neutrinos~\cite{Abbasi:2011jx}, a comparable constraint on the cross section of $\chi\chi\to \nu\bar{\nu}$ annihilations, which could take place in the core of Earth, is derived~\cite{Albuquerque:2011ma} for TeV range dark-matter masses.  Furthermore, upper limits on the annihilation cross sections of $\chi\chi\to \mu^+\mu^-$ and $\chi\chi\to \tau^+\tau^-$
channels are also obtained. It is interesting to compare these upper limits with the required annihilation cross sections for the same channels for explaining the PAMELA data on positron fraction excess
 \cite{Adriani:2011xv} and  Fermi-LAT $e^+ + e^-$ fluxes measurement \cite{Ackermann:2010ij}. As shown in Ref.~\cite{Abbasi:2011eq}, the IceCube upper limit 
on $\chi\chi\to \tau^+\tau^-$ annihilation cross section is comparable  to the required  $\chi\chi\to \tau^+\tau^-$ annihilation cross section for explaining PAMELA and Fermi-LAT data. On the other hand, the IceCube upper limit  on  $\chi\chi\to \mu^+\mu^-$ annihilation cross section  is still too high to test  the idea of using this mode to account for  PAMELA and Fermi-LAT data.  

The IceCube sensitivity on DM search is expected to improve with the data from all 86 strings analyzed. The analysis of DeepCore array data will further enhance the sensitivity. The DeepCore array \cite{Resconi:2009,Wiebusch:2009jf,DeYoung:2011ke} is located in
the deep center region of IceCube detector. This array consists of
$8$ densely instrumented strings plus the nearest standard IceCube
strings. The installation of DeepCore array significantly improves
the rejection of downward going atmospheric muons in IceCube and
lowers the threshold energy for detecting muon track or cascade
events to about $5$ GeV.  This muon rejection is crucial for IceCube to observe DM induced 
neutrino signature from galactic halo.  In fact, it has been pointed out that the parameter range 
for $\chi\chi\to \mu^+\mu^-$ channel preferred by PAMELA and Fermi-LAT data 
could  be stringently constrained
\cite{Spolyar:2009kx,Buckley:2009kw,Mandal:2009yk}(see also
discussions in Ref.~\cite{Covi:2009xn}) by the data from IceCube
detector augmented with DeepCore array. In this work, we shall however focus on the low-mass DM instead of DM with its mass in the range preferred
by PAMELA and Fermi-LAT.

We note that previous analyses on DeepCore sensitivity 
\cite{Mandal:2009yk,Erkoca:2010} have set the threshold energy at
$(40-50)$ GeV for both track and cascade events. However, to take the full advantage of DeepCore
array, it is desirable to include neutrino events in the energy range  $10 \ {\rm GeV}\leq
E_{\nu}\leq 50 \ {\rm GeV}$. We have initiated such a study for track events ~\cite{Lee:2012}. In this work,
we generalize the previous study to cascade events in IceCube DeepCore detector. 
Since we are only interested in low-mass DM, we only consider channels $\chi\chi\to b\bar{b}, \
\tau^+ \tau^-$, $\mu^+\mu^-$, and $\nu\bar{\nu}$ for annihilations and channels
$\chi\to b\bar{b}, \ \tau^+ \tau^-$ and $\mu^+\mu^-$, and  $\nu\bar{\nu}$ for decays.
The neutrino fluxes generated through DM annihilations or
decays into $t\bar{t}$, $W^+W^-$ and $ZZ$ final states are not included in this analysis.  

This paper is organized as follows. In Sec. II we
describe neutrino fluxes from DM annihilation/decay at
the galactic halo for different halo profiles.
In Sec. III we briefly describe our results on the background atmospheric neutrino  fluxes
taking into account neutrino oscillations.
In Sec. IV we present our results on the projected five year sensitivity of DeepCore array  on cascade events
induced by DM annihilations and decays. 
In addition, we shall also compare our results with
the up-to-date indirect detection limits from
Fermi-LAT gamma ray data and cosmic microwave background (CMB) observations. Specifically, we  shall compare the DeepCore sensitivities on 
DM annihilation cross section $\langle\sigma\upsilon\rangle$ and DM decay time with corresponding
constraints obtained from gamma ray observations \cite{Fermi:2012,Ackermann:2011}(see also analysis in Ref.~\cite{Cirelli:2010}) 
and those obtained from CMB anisotropy~\cite{Hutsi:2011vx,Galli:2011} based on one or both of the recent WMAP 7-year~\cite{Komatsu:2011} and ACT 2008~\cite{Fowler:2010} data.   
Discussions and conclusions are given in Sec. V.

\section{Neutrino Flux from Annihilations and Decays of Dark Matter in the Galactic Halo}

The differential neutrino flux from the galactic dark matter halo
for neutrino flavor $i$ can be written as~\cite{Hisano:2009}
\begin{equation}
\frac{\mbox{d}\Phi_{\nu_{i}}}{\mbox{d}E_{\nu_{i}}}=
\frac{\Delta\Omega}{4\pi}\frac{\langle\sigma\upsilon\rangle}{2m^{2}_{\chi}}
\left(\sum_F B_{F}\frac{\mbox{d}N^{F}_{\nu_{i}}}{\mbox{d}E}\right)
R_{\odot}\rho^{2}_{\odot}\times J_{2}(\Delta\Omega) \\
\label{eq1}
\end{equation}
for the case of annihilating DM, and
\begin{equation}
\frac{\mbox
{d}\Phi_{\nu_{i}}}{\mbox{d}E_{\nu_{i}}}=\frac{\Delta\Omega}{4\pi}\frac{1}{m_{\chi}\tau_{\chi}}
\left(\sum_F B_{F}\frac{\mbox{d}N^{F}_{\nu_{i}}}{\mbox{d}E}\right)
R_{\odot}\rho_{\odot}\times J_{1}(\Delta\Omega) \\
\label{eq2}
\end{equation}
for the case of decaying DM, where $R_{\odot} = 8.5~\textrm{kpc}$ is
the distance from the galactic center (GC) to the solar system,
$\rho_{\odot}=0.3~\textrm{GeV}/\textrm{cm}^{3}$ is the DM density in
the solar neighborhood, $m_{\chi}$ is the DM mass, $\tau_{\chi}$ is
the DM decay time and $\mbox{d}N^{F}_{\nu_{i}}/\mbox{d}E$ is the
neutrino spectrum per annihilation or decay for a given annihilation
or decay channel $F$ with a corresponding branching fraction
$B_{F}$. For $\chi\chi\to \nu\overline{\nu}$ channel, we assume
that two neutrinos are produced per annihilated DM pair and
all neutrino flavors are equally populated. Thus the neutrino spectrum
per flavor is a monochromatic line with
$\mbox{d}N_{\nu}/\mbox{d}E=\frac{2}{3}\delta(E-m_{\chi})$.
On the other hand, the neutrino spectrum per flavor for $\chi\to \nu\overline{\nu}$ channel
is $\mbox{d}N_{\nu}/\mbox{d}E=\frac{2}{3}\delta(E-\frac{m_{\chi}}{2})$.
The neutrino spectra $\mbox{d}N^{F}_{\nu_{i}}/\mbox{d}E$
for other channels are summarized in Refs~\cite{Erkoca:2010,Reno:2010}. The
quantity $\langle\sigma\upsilon\rangle$ is the thermally averaged
annihilation cross section, which can be written as
\begin{equation}
\langle\sigma\upsilon\rangle = B \langle\sigma\upsilon\rangle_{0},           \\
\label{eq3}
\end{equation}
with a boost factor $B$ \cite{boost_factor}. We set
$\langle\sigma\upsilon\rangle_{0}=3\times10^{-26}~\textrm{cm}^{3}\textrm{s}^{-1}$,
which is the typical annihilation cross section for the present dark
matter abundance under the standard thermal relic scenario
\cite{Jungman:1996}. We treat the boost factor $B$ as a
phenomenological parameter. The dimensionless quantity
$J_{n}(\Delta\Omega)$ is the DM distribution integrated over the
line-of-sight (l.o.s) and averaged over a solid angle
$\Delta\Omega=2\pi(1 - \textrm{cos}\psi_{\textrm{max}} )$, i.e.,
\begin{equation}
J_{n}(\Delta\Omega) =
\frac{1}{\Delta\Omega}\int_{\Delta\Omega}\mbox{d}\Omega\int_{\rm{l.o.s}}\frac{\mbox{d}l}{R_{\odot}}
\left(\frac{\rho(r(l,\psi))}{\rho_{\odot}}\right)^{n},           \\
\label{eq4}
\end{equation}
\\
where $\rho$ is the DM density at a specific location described by
the coordinate $(l,\psi)$ with $l$ the distance from the Earth to DM
and $\psi$ the direction of DM viewed from the Earth with $\psi=0$
corresponding to the direction of GC. The distance $r\equiv
\sqrt{R^{2}_{\odot} + l^{2} - 2R_{\odot}l\textrm{cos}\psi }$ is the
distance from GC to DM. We note that the above definition of $J_n$ include the constant  factors ${\Delta\Omega}$, $R_{\odot}$ and $\rho_{\odot}$
in the denominator. This definition differs from that adopted in papers by IceCube and 
Fermi-LAT collaborations (see, for example, Refs.~\cite{Abbasi:2011eq} and \cite{Ackermann:2011}) where the above-mentioned constant factors 
are not included.  We computed the values of $J_{n}(\Delta\Omega)$ with DarkSUSY \cite{Gondolo:2004sc}. For the galactic DM distribution,
we consider Navarro-Frenk-White (NFW) ~\cite{NFW},
Einasto~\cite{Einasto1,Einasto2,Diemand:2008} and Isothermal profiles~\cite{Isothermal}.
The functional forms of these profiles are given by
\begin{eqnarray}
\label{NFW}
\rho_{\textrm{NFW}}(r) = \rho_{s}\frac{R_{s}}{r}\left(1+\frac{r}{R_{s}}\right)^{-2},
\end{eqnarray}
\begin{eqnarray}
\label{Einasto}
\rho_{\textrm{Ein}}(r) = \rho_{s}\times\textrm{exp}\left\{-\frac{2}{\alpha}\left[\left(\frac{r}{R_{s}}\right)^{\alpha}-1\right]\right\}, \,\alpha = 0.17,
\end{eqnarray}
\begin{eqnarray}
\label{Isothermal}
\rho_{\textrm{Iso}}(r) = \frac{\rho_{s}}{1+(r/R_{s})^{2}},
\end{eqnarray}
with values of  $R_s$ and $\rho_s$ given in Table I.
In all cases we impose the normalization $\rho_{\odot}=0.3~\textrm{GeV}/\textrm{cm}^{3}$, which is at $r=8.5$ kpc.
\begin{table}
\caption{Parameters of the density profiles for DM halo.}
\begin{center}
\begin{ruledtabular}
\begin{tabular}{ccc}
DM halo model & $R_{s}$ in kpc & $\rho_{s}$ in $\textrm{GeV}/\textrm{cm}^{3}$
\\ \hline \\
NFW  & 20.0  & 0.260      \\
Einasto  & 21.5  & 0.0538       \\
Isothermal & 3.50  & 2.07      \\
\end{tabular}
\end{ruledtabular}
\end{center}
\end{table}
The comparison of different DM halo models is shown in Fig.~1.

Neutrinos are significantly mixed through oscillations when they
travel a vast distance across the galaxy. The observed flavor ratio of DM induced neutrinos is related to the flavor ratio at the source through the probability matrix $P_{\alpha\beta}\equiv P(\nu_{\beta}\to \nu_{\alpha})$ ~\cite{Learned:1994wg,AB,Lai}. In particular, the exact form of $P_{\alpha\beta}$ in terms of mixing angles $\theta_{ij}$ and CP phase $\delta$ is given in Ref.~\cite{Lai}.
The mixing angles $\theta_{23}$ and $\theta_{12}$ has been well measured while newest results
for $\theta_{13}$ from accelerator~\cite{Abe:2011sj} and reactor experiments~\cite{Abe:2011fz, An:2012eh, Ahn:2012nd} are also available.  We take
$\sin^2 \theta_{23}=0.386$, $\sin^2 \theta_{12}=0.307$, $\sin^2 \theta_{13}=0.0241$ and $\delta=1.08\pi$, which are best fit values of neutrino mixing parameters from a recent global fitting~\cite{Fogli:2012ua} in the case of normal mass hierarchy. Therefore,
\begin{eqnarray}
\Phi_{\nu_{e}}& =&
0.55\Phi_{\nu_{e}}^{0}+0.24\Phi_{\nu_{\mu}}^{0}+0.21
\Phi_{\nu_{\tau}}^{0},\nonumber \\
\Phi_{\nu_{\mu}}& =&
0.24\Phi_{\nu_{e}}^{0}+0.40\Phi_{\nu_{\mu}}^{0}+0.35
\Phi_{\nu_{\tau}}^{0},\nonumber\\
\Phi_{\nu_{\tau}}& =&
0.21\Phi_{\nu_{e}}^{0}+0.35\Phi_{\nu_{\mu}}^{0}+0.44
\Phi_{\nu_{\tau}}^{0},
\end{eqnarray}
where $\Phi_{\nu_i}^0$ and $\Phi_{\nu_{\nu_i}}$ are neutrino fluxes at the source and on the earth respectively.
While the best fit neutrino mixing parameters differ slightly in the case of inverted mass hierarchy, they do not produce noticeable change in the above relation. 
\section{ATMOSPHERIC NEUTRINO FLUXES }

We follow the approaches in ~\cite{Gaisser:2001sd, Lee:2006, Lee:2012}
to compute the intrinsic atmospheric neutrino background fluxes.
The $\nu_{\mu}$ flux arising from $\pi$ decays reads
\begin{eqnarray}
\label{atm-nu}
 \frac{\mbox{d}^2N^{\pi}_{\nu_{\mu}}(E,\xi,X)}{\mbox{d}E\mbox{d}X}&=&\int_E^{\infty}
 \mbox{d}E_N\int_E^{E_N}
 \mbox{d}E_{\pi}\frac{\Theta(E_{\pi}-\frac{E}{1-\gamma_{\pi}})}{d_{\pi}E_{\pi}(1-\gamma_{\pi})}
 \nonumber \\
 & &\times \int_0^X
 \frac{\mbox{d}X'}{\lambda_N}P_{\pi}(E_{\pi},X,X')
 \frac{1}{E_{\pi}}F_{N\pi}(E_{\pi},E_N)\nonumber \\
 & & \times \exp \left(-\frac{X'}{\Lambda_N}\right)\phi_N(E_N),
\end{eqnarray}
where $E$ is the neutrino energy, $X$ is the slant depth in units of
g/cm$^2$,  $\xi$ is the zenith angle in the direction of incident
cosmic-ray nucleons, $r_{\pi}=m_{\mu}^2/m_{\pi}^2$, $d_{\pi}$ is the
pion decay length in units of g/cm$^2$, $\lambda_N$ is the nucleon
interaction length while $\Lambda_N$ is the corresponding nucleon
attenuation length, and $\phi_N(E_N)$ is the primary cosmic-ray spectrum.
We have $\phi_N(E_N)=\sum_{A}A\phi_A(E_N)$ with $A$
the atomic number of each nucleus. The spectrum of each cosmic-ray
component is parametrized by \cite{Gaisser:2002jj,Honda:2004}
\begin{eqnarray}
\label{phi n}
\phi_A(E_N) = K\times(E_N + b\,\textrm{exp}[- c
\sqrt{E_N}])^{- \alpha},
\end{eqnarray}
in units of m$^{-2}$s$^{-1}$sr$^{-1}$GeV$^{-1}$. The fitting
parameters $\alpha,K,b,c$ depend on the type of nucleus. They are
tabulated in Ref.~\cite{Honda:2004}. The function
$P_{\pi}(E_{\pi},X,X')$ is the probability that a charged pion produced
at the slant depth $X'$ survives to the depth $X$ ($> X'$) ~\cite{Lipari:1993hd}.
$F_{N\pi}(E_{\pi},E_N)$ is the normalized inclusive cross section
for $N+{\rm air}\to \pi^{\pm}+Y$, and is given in Ref.~\cite{Gaisser:2001sd}.
The kaon contribution to the atmospheric $\nu_{\mu}$ flux has the same
form as Eq.~(\ref{atm-nu}) with an inclusion of the branching ratio
$B(K\to \mu\nu)=0.635$ and appropriate replacements in kinematic factors
as well as in the normalized inclusive cross section. 
The three-body muon decay contribution
to the atmospheric $\nu_{\mu}$ flux is also included. The details are discussed  in Ref.~\cite{Lee:2006}.
After summing the two-body and three-body decay contributions, we obtain
the total intrinsic atmospheric muon neutrino flux. From the Fig.~1 of Ref.~\cite{Lee:2012},
we note that angle-averaged atmospheric muon neutrino flux obtained by our
calculation and that obtained by Honda et al. \cite{Honda:2007} both
agree well with AMANDA-II results \cite{Amanda:2010}.

\begin{figure}
  \centering
  \includegraphics[width=0.85\textwidth]{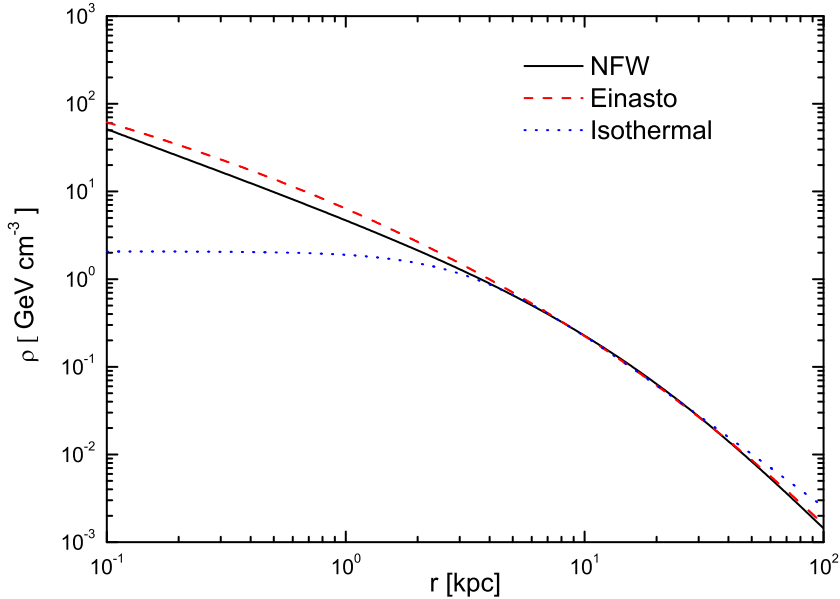}
  \vskip 0.5cm
  \caption{ Comparison of the dark matter density distribution,
  $\rho(r)$, as a function of distance from the Galactic Center as
  described by the NFW, Einasto and Isothermal halo
  models.}
  \label{fig:1}
\end{figure}

The intrinsic atmospheric $\nu_{\tau}$ flux due to $D_s$ decays can be obtained
by solving cascade equations ~\cite{Gaisser:1992vg,Lee:2006}. One obtains
\begin{equation}
\label{atm-tau}
 \frac{\mbox{d}^2N_{\nu_{\tau}}(E,X)}{\mbox{d}E\mbox{d}X}=
 \frac{Z_{ND_s}Z_{D_s\nu_{\tau}}}{1-Z_{NN}}\cdot
 \frac{\exp(-X/\Lambda_N)\phi_N(E_N)}{\Lambda_N},
 \end{equation}
where $Z_{NN}\equiv 1-\lambda_N/\Lambda_N$, $Z_{ND_s}$ and $Z_{D_s\nu_{\tau}}$
are the Z moments defined in our previous work~\cite{Lee:2012}. Finally, the atmospheric $\nu_{\mu}$ flux taking into account the
neutrino oscillation effect is given by
\begin{eqnarray}
\label{oscillate1}
\frac{\mbox{d}\bar{N}_{\nu_{\mu}}(E,\xi)}{\mbox{d}E}&=&\int
{\mbox d}X\left[
\frac{\mbox{d}^2N_{\nu_{\tau}}}{\mbox{d}E\mbox{d}X}\cdot
P_{\nu_{\tau}\to \nu_{\mu}}\right.\nonumber \\
&&\left.+\frac{\mbox{d}^2N_{\nu_{\mu}}}{\mbox{d}E\mbox{d}X}
\cdot \left(1-P_{\nu_{\mu}\to
\nu_{\tau}}\right)\right],
\end{eqnarray}
where $P_{\nu_{\alpha}\to \nu_{\beta}}$ is the $\nu_{\alpha}\to \nu_{\beta}$ oscillation probability. Sub-leading contributions to atmospheric $\nu_{\mu}$ flux arising from $\nu_{\mu}\to \nu_e$
and $\nu_e\to \nu_{\mu}$ oscillations are not included in the above equation. 
We can write down the atmospheric $\nu_{\tau}$ flux 
in the similar way.

\section{RESULTS}

In IceCube DeepCore, the track event rate for contained muons is given by
\begin{eqnarray}
\label{muevent}
\Gamma_{\mu}=\int_{E_\mu^{\rm{th}}}^{E_{\rm{max}}}{\mbox
d}E_\mu
\int_{E_\mu}^{E_{\rm{max}}}{\mbox d}E_{\nu_{\mu}} N_{A} \rho_{\rm{ice}} V_{\rm{tr}}
\times\frac{d\Phi_{\nu_{\mu}}}{dE_{\nu_{\mu}}}\cdot \frac{d\sigma_{\nu N}^{\rm{CC}}(E_{\nu_{\mu}},E_{\mu})}{dE_{\mu}}
+(\nu \rightarrow \overline{\nu}),
\end{eqnarray}
while the cascade event rate is given by
\begin{eqnarray}
\label{casevent}
\Gamma_{\rm{casc}}=\int_{E_{sh}^{\rm{th}}}^{E_{\rm{max}}}{\mbox
d}E_{sh}
\int_{E_{sh}}^{E_{\rm{max}}}{\mbox d}E_{\nu} N_{A} \rho_{\rm{ice}} V_{\rm{casc}}
\times\frac{d\Phi_{\nu}}{dE_{\nu}}\cdot \frac{d\sigma_{\nu N}(E_{\nu},E_{sh})}{dE_{sh}}
+(\nu \rightarrow \overline{\nu}),
\end{eqnarray}
where $\rho_{\textrm{ice}} = 0.9\, \textrm{g}\,\textrm{cm}^{-3}$ is
the density of ice, $N_{A} = 6.022\times10^{23}\,\textrm{g}^{-1}$ is
Avogadro's number, $V_{\textrm{tr}}\approx0.04\,\textrm{km}^{3}$ is
the effective volume of IceCube DeepCore array for muon track events~\cite{Resconi:2009,Mandal:2009yk}
and $V_{\rm{casc}}\approx0.02\,\rm{km}^{3}$ is that for cascade events
\cite{Mandal:2009yk,McDonald:2003xn}, $d\Phi_{\nu}/dE_{\nu}$ is the
neutrino flux arrived at IceCube, which is the sum of DM induced flux and the background atmospheric neutrino flux, $E_{\textrm{max}}$ is taken as
$m_{\chi}$ for annihilation and $m_{\chi}/2$ for decay,
$E_\mu^{\textrm{th}}$ and $E_{sh}^{\rm{th}}$ are the threshold energies for track events
and cascade events respectively,  $d\sigma_{\nu N}^{\textrm{CC}}/dE_{\mu}$ is
the differential cross section of neutrino-nucleon charged-current scattering, and $d\sigma_{\nu N}/dE_{sh}$ is the differential cross section for
showers produced by neutrino-nucleon charged-current and neutral-current scatterings. In this work, we use differential cross sections $d\sigma_{\nu
N}^{\textrm{CC}}/dE_{\mu}$ and $d\sigma_{\nu N}/dE_{sh}$ given by Ref.~
\cite{Gandhi:1996} with CTEQ6 parton distribution functions.
The atmospheric part of $d\Phi_{\nu_e}/dE_{\nu_e}$ is taken from Ref.~\cite{Honda:2007}.
We also set $E_\mu^{\textrm{th}}=E_{sh}^{\textrm{th}}=10$ GeV. It should be noted that the value for $V_{\textrm{tr}}$ is an average effective volume based on the 
energy dependent $V_{\textrm{tr}}$ discussed in Ref.~\cite{Resconi:2009}, while the value for $V_{\rm{casc}}$ is just the instrumental volume of the
DeepCore detector. The updated effective volumes of DeepCore detector for track and cascade events are available in Ref.~\cite{Collaboration:2011ym}. While we 
adopt constant effective volumes for evaluating DeepCore sensitivities,  we shall also estimate how much the updated effective volumes affect our results.      

\begin{figure}
  \centering
  \includegraphics[width=0.85\textwidth]{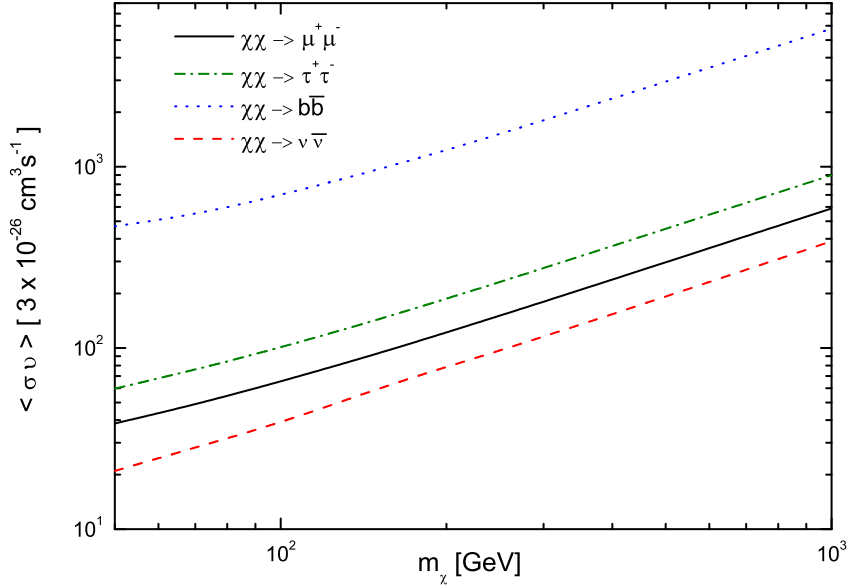}
  \vskip 0.5cm
  \caption{ The dotted line, dot-dashed line, solid line, and dashed line
  are the expected DeepCore sensitivities to DM annihilation cross section with the detection of cascade events from
  $\chi\chi\rightarrow b\overline{b}$,
  $\chi\chi\rightarrow\tau^+\tau^-$,
  $\chi\chi\rightarrow \mu^{+}\mu^{-}$, and $\chi\chi\rightarrow \nu\overline{\nu}$ channels, respectively. We adopt NFW profile for obtaining the results in this figure and those in the subsequent figures except 
Fig.~4. }
  \label{fig:2}
\end{figure}

As mentioned earlier, we consider neutrino fluxes generated through the
annihilation channels $\chi\chi\to b\bar{b}, \ \tau^+ \tau^-$,
$\mu^+\mu^-$ and $\nu\overline{\nu}$, and the decay channels $\chi\to b\bar{b}, \ \tau^+
\tau^-$, $\mu^+\mu^-$ and $\nu\overline{\nu}$. By computing the cascade event rates,
we present in Fig.~2 the required DM annihilation cross section as a
function of $m_{\chi}$ such that the neutrino signature
from DM annihilations can be detected at the $2\sigma$ significance
in five years. The $2\sigma$ statistical significance is defined as 
\begin{eqnarray}
\label{significance}
\frac{N_s}{\sqrt{N_s+N_b}}=2,
\end{eqnarray}
where $N_s$ and $N_b$ are numbers of signal and background events, respectively. 
In Fig.~2, we take NFW  profile for DM density distribution in the galactic halo, the shower threshold energy  $E_{sh}^{\textrm{th}}=
10~\textrm{GeV}$ and the cone half-angle
$\psi_{\textrm{max}}=50^{\circ}$. Non-detection of DM neutrino signature would then exclude
the parameter region above the curve at the $2\sigma$ level. We have
presented results corresponding to different annihilation channels.
It is seen that the required annihilation cross section for
$2\sigma$ detection significance is smallest for
$\chi\chi\rightarrow\nu\overline{\nu}$ channel and largest for the
channel $\chi\chi\rightarrow b\overline{b}$. 

At this juncture, it is desirable to estimate  effects of the updated effective volume on our sensitivity calculations. 
The energy dependent  effective volume $V_{\rm{casc}}(E)$ as given in Ref.~\cite{Collaboration:2011ym} is roughly 3 times smaller than the value $0.02\,\rm{km}^{3}$ adopted in our calculation
for $E_{\nu}=10$ GeV. On the other hand,     
$V_{\rm{casc}}(E)$ increases monotonically with energy with $V_{\rm{casc}}(E)$ greater than $0.02\,\rm{km}^{3}$ for $E_{\nu}>40$ GeV. For the annihilation process $\chi\chi\to \nu\bar{\nu}$,
one has $E_{\nu}=m_{\chi}$. Hence the DeepCore sensitivity to this process should be better than that presented in Fig.~2, which starts from $m_{\chi}= 50$ GeV. For $\chi\chi\to \mu^+\mu^-$ mode, 
$m_{\chi}=50$ GeV corresponds to $E_{\nu}\simeq 20$ GeV. At this energy, the value  $V_{\rm{casc}}=0.02\,\rm{km}^{3}$ overestimates the effective volume by about  a factor of 2, thus the DeepCore sensitivity to $\langle\sigma(\chi\chi\to \mu^+\mu^-)\upsilon\rangle$ should be corrected by a factor of $\sqrt{2}\approx 1.4$. The correction factor gradually reduces to $1$ 
as $m_{\chi}$ approaches to about $120$ GeV, which corresponds to $E_{\nu}= 40$ GeV. For $m_{\chi}> 120$ GeV, the DeepCore sensitivity calculated with $V_{\rm{casc}}(E)$ is better than that presented in Fig.~2. The correction factor for $\chi\chi\to \tau^+\tau^-$ is similar to that for $\chi\chi\to \mu^+\mu^-$. The hadronic mode $\chi\chi\to b\bar{b}$ requires different correction factor for the same 
$m_{\chi}$. However we shall not address such a correction here since DeepCore detector is relatively insensitive to $\chi\chi\to b\bar{b}$.    
\begin{figure}
  \centering
  \includegraphics[width=0.85\textwidth]{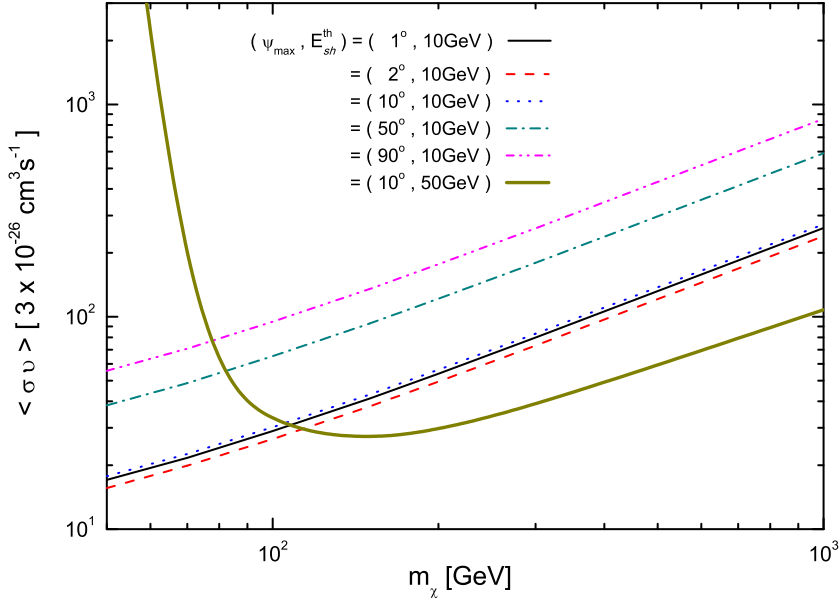}
  \vskip 0.5cm
  \caption{The required DM annihilation cross section
  $(\chi\chi\rightarrow\mu^{+}\mu^{-})$ as a function of $m_{\chi}$
  such that the cascade events induced by neutrinos from DM annihilations can be detected at the $2\sigma$
  significance in five years. Results corresponding to different $\psi_{\rm max}$ are
  presented. For comparison, we also show the result with $E_{sh}^{\rm th}=50$ GeV
  and $\psi_{\textrm{max}}=10^{\circ}$ \cite{Erkoca:2010}.
  }
  \label{fig:3}
\end{figure}

Next, we show how the DeepCore sensitivity on DM annihilation cross section
varies with the chosen cone half-angle and threshold energy for NFW DM density profile.
Here we take $\chi\chi\to \mu^+\mu^-$ channel for illustration. Fig.~3 shows the required
DM annihilation cross section $\langle\sigma(\chi\chi\to \mu^+\mu^-)\upsilon\rangle$ for a $2\sigma$
detection in five years for different cone half-angle $\psi_{\textrm{max}}$.
It is seen that
the sensitivity is improved as $\psi_{\textrm{max}}$ increases from $1^{\circ}$
to $2^{\circ}$ while it turns weaker as $\psi_{\textrm{max}}$ increases further.
In the latter case, the signal increases slower
than the background does. We should point out that the choice of  $\psi_{\textrm{max}}$ depends on the 
angular resolution of the experiment. The current angular resolution in IceCube for cascade events is $50^{\circ}$. However,  an improvement on such a resolution is expected~\cite{Middell}.
In fact, new reconstruction method which can achieve a $5^{\circ}$ angular resolution for cascade events in large-scale neutrino telescopes has been proposed~\cite{Auer}.   
Hence results shown in Fig.~3 with $\psi_{\textrm{max}}\ge 10^{\circ}$ can be realized in the near future.
In this figure, we also show the
result for a higher threshold energy $E_{sh}^{\textrm{th}}=
50~\textrm{GeV}$ with a cone half-angle
$\psi_{\textrm{max}}=10^{\circ}$ for comparison. This result is
taken from Ref.~\cite{Erkoca:2010} where
$\psi_{\textrm{max}}=10^{\circ}$ is identified as the most optimal
cone half-angle for constraining DM annihilation cross section at
that threshold energy. 
For large $m_{\chi}$, lowering $E_{sh}^{\textrm{th}}$ from $50$ GeV to $10$ GeV 
does not affect much the signal rate while increases significantly the atmospheric background event rate. Hence, the sensitivity on DM annihilation cross
section becomes worse by choosing $E_{sh}^{\textrm{th}}=
10~\textrm{GeV}$. One expects the situation turns opposite for $m_{\chi}$ approaching to the threshold energy. In fact,
for $m_{\chi}<100$ GeV, one can see that the sensitivity on DM
annihilation cross section obtained with $E_{sh}^{\textrm{th}}=
10~\textrm{GeV}$ is always better than that obtained with
$E_{sh}^{\textrm{th}}= 50~\textrm{GeV}$. We note that the DeepCore sensitivities on
other annihilation channels have similar cone half-angle
and threshold energy dependencies. From the above discussions, we observe that, in contrast to the main concern of this article,  raising the shower threshold energy gains sensitivity for probing heavier DM. In fact, as the threshold energy 
approaches to $100$ GeV, one enters into the operative energy range of the full IceCube array such that the effective volume of the detector increases more rapidly with the threshold energy 
than the case with DeepCore detector alone~\cite{Collaboration:2011ym}.  The same situation holds for probing DM annihilation with track events.

\begin{figure}
  \centering
  \includegraphics[width=0.85\textwidth]{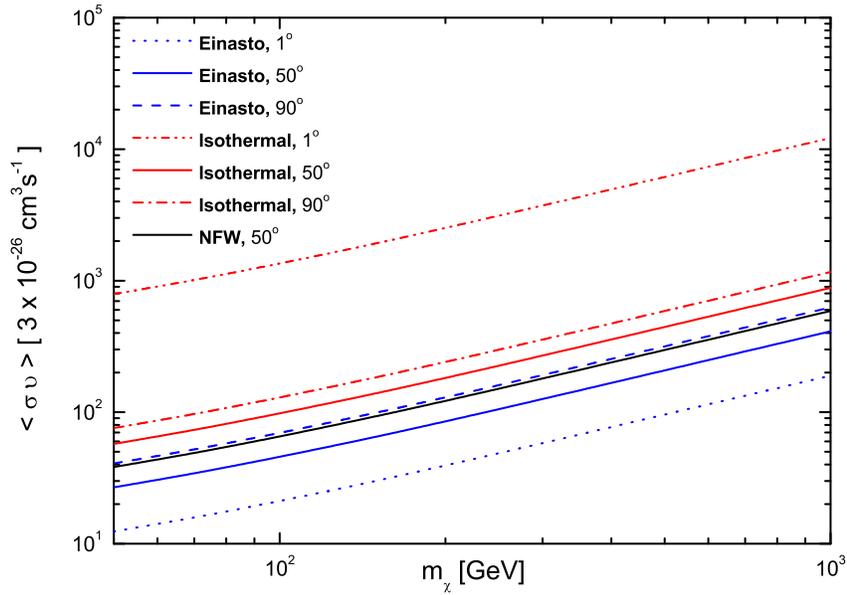}
  \vskip 0.5cm
  \caption{ The expected DeepCore  sensitivities corresponding to different $\psi_{\textrm{max}}$
  for Einasto ~\cite{Einasto1,Einasto2} , Isothermal ~\cite{Isothermal}and NFW ~\cite{NFW} DM density profiles.}
  \label{fig:4}
\end{figure}

After discussing how the DeepCore sensitivities on DM annihilation cross section
vary with the chosen cone half-angle and threshold energy for NFW profile, we study the variation 
of DeepCore sensitivities on $\chi\chi\to \mu^+\mu^-$
with $\psi_{\textrm{max}}$ for different DM density profiles. First, we present in
Fig.~4 our expected DeepCore sensitivities corresponding to different $\psi_{\textrm{max}}$
for Einasto DM density profile. Similar to NFW profile, Einasto DM density distribution  also has a cusp
in the central DM region. We refer to this class of DM
density profiles as the cusped profile. Therefore the DeepCore sensitivity on $\langle\sigma(\chi\chi\to \mu^+\mu^-)\upsilon\rangle$ with Einasto profile becomes poorer as
$\psi_{\textrm{max}}$ increases. In addition, we can see from Fig.~1 that the DM density of
Einasto profile is higher than that of NFW profile between 0.1 kpc and 5 kpc from the GC, and both density profiles are almost
identical beyond this range of distances. Hence the DeepCore sensitivity on $\langle\sigma(\chi\chi\to \mu^+\mu^-)\upsilon\rangle$ with Einasto profile is better than that with NFW profile for
$\psi_{\textrm{max}}=50^{\circ}$. Next, we also present in Fig.~4  the expected DeepCore sensitivities
for different $\psi_{\textrm{max}}$ with Isothermal DM density profile. Because there is a core in the central DM density distribution of Isothermal profile,
we refer to this class of profile as the cored profile. We note that the sensitivity to DM annihilation cross section with
Isothermal profile improves as $\psi_{\textrm{max}}$ increases from $1^{\circ}$ to $50^{\circ}$ while the sensitivity with Einasto profile behaves oppositely.  
This is because that the DM density
distribution of Isothermal profile maintains flat for a much longer distance
from the GC as compared to the cusped profile. However,  as  $\psi_{\textrm{max}}$ increases further, the DeepCore sensitivity to DM annihilation cross section  
becomes poorer even for Isothermal profile since the factor
$J_{2}(\Delta\Omega)\Delta\Omega$ is proportional to the square of DM density. 
Finally, DeepCore sensitivities to DM annihilation cross section with Isothermal profile are poorer than those with cusped profiles
for the same $\psi_{\textrm{max}}$. This results from the fact that DM densities of cusped profiles around GC are
much larger than that of Isothermal profile in the same region.

\begin{figure}
  \centering
  \includegraphics[width=0.85\textwidth]{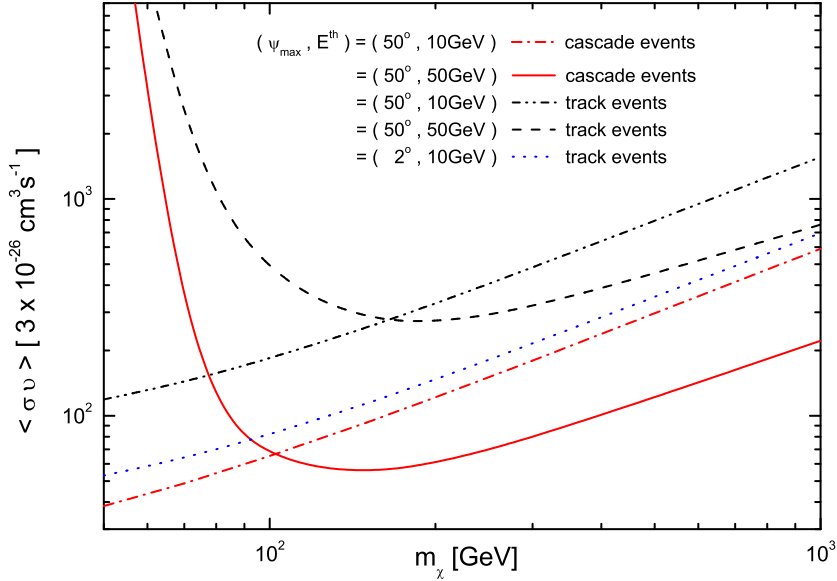}
  \vskip 0.5cm
  \caption{ The required DM annihilation cross section
  $(\chi\chi\rightarrow\mu^{+}\mu^{-})$ as a function of $m_{\chi}$
  such that the neutrino signature from DM annihilations can be detected at the $2\sigma$
  significance in five years for track and cascade events.}
  \label{fig:5}
\end{figure}

Having discussed the effect of DM density profiles on the derived DeepCore sensitivities,  we present in Fig.~5 the comparison of DeepCore  sensitivities to $\langle\sigma(\chi\chi\to \mu^+\mu^-)\upsilon\rangle$ obtained by measuring cascade events and track events, respectively. We note that NFW profile is adopted for results in this figure as well as results presented  in the remaining
figures of this article. 
We point out  that $\nu_{\mu}$ is the dominant flavor of atmospheric neutrinos
above few tens of GeV and the neutrino-nucleon cross section are almost the same for all flavors.
Therefore, comparing with the result of track events, the signal to background ratio is enhanced
in cascade events because $\nu_{\mu}$ only produces the cascade events through the neutral-current
interaction, which is lower in cross section than that of charged-current interaction. Hence
cascade events in general provide better sensitivities to DM annihilation cross section than those provided
by track events.  We like to point out that the comparison between cascade and track events in Fig.~5 is based upon the current angular resolution of IceCube detector.
If the angular resolution for cascade events is improved in the future, the advantage of measuring cascade events
would be even more significant as one can see from Fig.~3.

\begin{figure}
  \centering
  \includegraphics[width=0.85\textwidth]{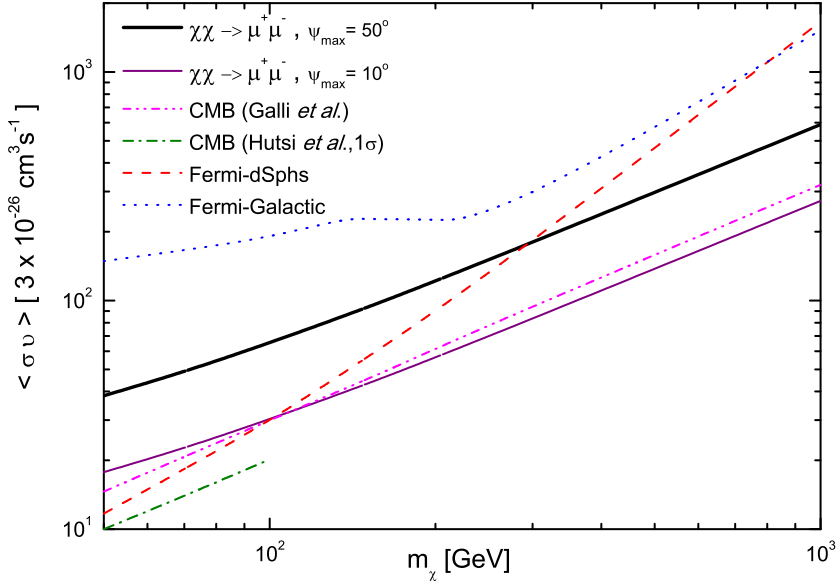}
  \vskip 0.5cm
  \caption{ Cross section limits on DM annihilation channel $\chi\chi\to \mu^+\mu^-$.
  The dot-dashed line is the $1\sigma$ upper bound on annihilation cross section for low-mass WIMP~\cite{Hutsi:2011vx} obtained by WMAP7 data.
 The dot-dot-dashed line is the CMB constraint obtained by using WMAP7 + ACT data at $95\%$ C.L. ~\cite{Galli:2011}.
  The dashed line is the dSphs constraint at $95\%$ C.L. ~\cite{Ackermann:2011}.
  The dotted line is the constraint due to Fermi-LAT observations on the region $|b| > 10^{\circ}$ plus a $20^{\circ}\times20^{\circ}$ square region centered
  at  GC, assuming the NFW profile~\cite{Fermi:2012}. The thick and thin solid lines are the expected $2\sigma$ sensitivities of DeepCore detector with
  $(E_{sh}^{\textrm{th}}, \psi_{\textrm{max}})=(10~\textrm{GeV}, 50^{\circ})$ and  $(E_{sh}^{\textrm{th}}, \psi_{\textrm{max}})=(10~\textrm{GeV}, 10^{\circ})$, respectively.}
  \label{fig:6}
\end{figure}

It is interesting and essential to compare our results with constraints obtained from gamma-ray astronomy and cosmology.
Fermi large area telescope (Fermi-LAT) ~\cite{Fermi:2009} is a pair-conversion telescope that explores the
gamma-ray sky in the 20 MeV to 300 GeV range with unprecedented sensitivity. In a recent work, 
Fermi-LAT collaboration derive constraints on WIMP annihilation
or decay into various final states which produce a continuous photon spectrum~\cite{Fermi:2012}.
These constraints are based upon the measured inclusive
photon intensity spectrum from 4.8 GeV to 264 GeV obtained from two years of Fermi-LAT data over the region
$|b| > 10^{\circ}$ plus a $20^{\circ}\times20^{\circ}$  square region centered at GC with point
sources removed. 
In Fig.~6, the dotted line is cross section upper limit on  DM annihilation channel $\chi\chi\to \mu^+\mu^-$ 
from the diffuse gamma-ray spectrum with NFW profile. It is taken from Ref.~\cite{Fermi:2012} 
with a rescaling factor $(4/3)^{2}$ applied since we have adopted  a local density of $\rho_{\odot}=0.3~\textrm{GeV}/\textrm{cm}^{3}$ while Fermi-LAT analysis uses  $\rho_{\odot}=0.4~\textrm{GeV}/\textrm{cm}^{3}$.
Our expected $2\sigma$ sensitivities on $\chi\chi\to \mu^+\mu^-$ annihilation cross section
by the DeepCore detector with $\psi_{\textrm{max}}=50^{\circ}$  and $\psi_{\textrm{max}}=10^{\circ}$  are plotted for comparison.
We can see that our expected $2\sigma$ sensitivity with
$\psi_{\textrm{max}}=50^{\circ}$ is slightly stronger than this Fermi-LAT constraint, even if we calculate the sensitivity with   
$V_{\rm{casc}}(E)$ given by Ref.~\cite{Collaboration:2011ym}.
Furthermore,  the expected $2\sigma$ sensitivity with
$\psi_{\textrm{max}}=10^{\circ}$ in the DeepCore detector is almost an order of magnitude stronger than this Fermi-LAT constraint for small $m_{\chi}$. 
\begin{figure}
  \centering
  \includegraphics[width=0.85\textwidth]{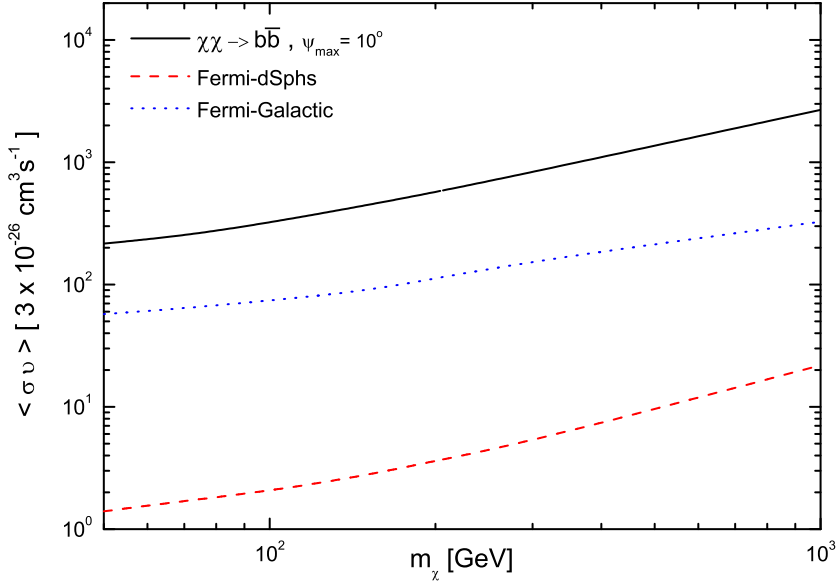}
  \vskip 0.5cm
  \caption{ Cross section limits on the DM annihilation channel $\chi\chi\to b\bar{b}$.
  The dashed line is the dSphs constraint at $95\%$ C.L. ~\cite{Ackermann:2011}.
  The dotted line is the constraint due to Fermi-LAT observations for the region $|b| > 10^{\circ}$ plus a $20^{\circ}\times20^{\circ}$ square region centered
  at the GC, assuming the NFW profile ~\cite{Fermi:2012}. The solid line is the expected $2\sigma$ sensitivity by DeepCore detector with
  $E_{sh}^{\textrm{th}}=10~\textrm{GeV}$ and $\psi_{\textrm{max}}=10^{\circ}$, which is weaker than both Fermi-LAT constraints. The DeepCore sensitivity with  $\psi_{\textrm{max}}=50^{\circ}$ is not shown since 
it is even less competitive.}
  \label{fig:7}
\end{figure}

Constraints on DM annihilation cross section were also obtained from
cosmology and gamma-ray observations on dwarf spheroidal satellite galaxies (dSphs) of the Milky Way.
Energy injection from DM annihilation at redshift $100\lesssim z \lesssim1000$
affects the CMB anisotropy. This is because the injected
energy can ionize the thermal gas and modify the standard recombination history
of the universe. The updated CMB constraints on DM annihilation cross sections
are derived in Refs.~\cite{Hutsi:2011vx} and \cite{Galli:2011}, where the former work is based on the recent WMAP 7-year data~\cite{Komatsu:2011} while 
the latter one combines WMAP 7-year and  ACT 2008~\cite{Fowler:2010} data. We present
the above two  CMB constraints on $\chi\chi\to \mu^+\mu^-$ annihilation cross section in Fig.~6.
One can see that the thick solid line is higher than the dot-dot-dashed line ($95\%$ C.L.) by roughly a factor of 2.
Thus the expected sensitivity of IceCube DeepCore detector with $\psi_{\textrm{max}}=50^{\circ}$ 
 is slightly weaker than the CMB constraint. 
However, the DeepCore sensitivity is comparable to the CMB constraint with $\psi_{\textrm{max}}=10^{\circ}$.
It should be noted that the DM annihilation cross section could be velocity dependent. Hence,  a model dependent extrapolation on DM annihilation cross section might be required to compare the constraint on $\langle \sigma\upsilon \rangle$ at  redshift $100\lesssim z \lesssim1000$ to that at the present day universe \cite{Hisano:2011}.

\begin{figure}
  \centering
  \includegraphics[width=0.85\textwidth]{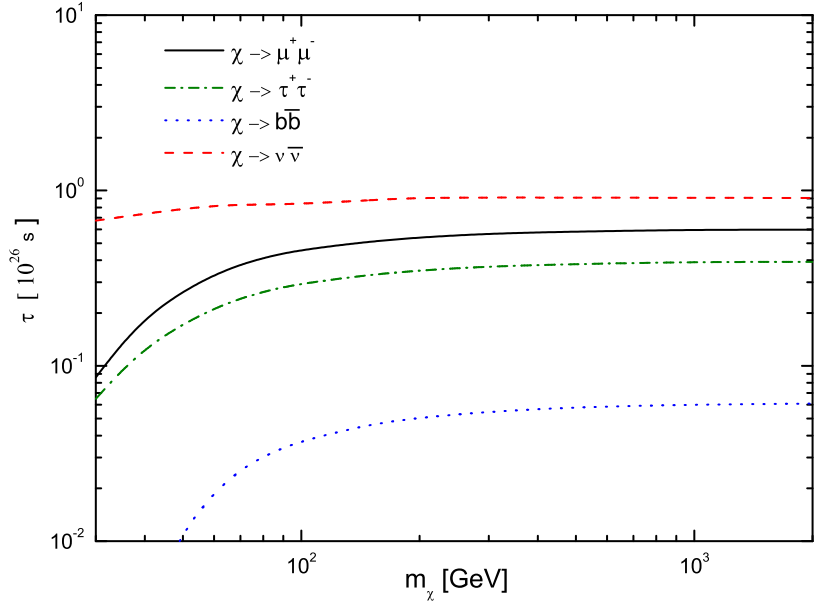}
  \vskip 0.5cm
  \caption{ The dotted line, the dot-dashed line, the solid line, and dashed line
  are the expected DeepCore sensitivities to DM decay time through cascade events  from decay channels
  $\chi\rightarrow b\overline{b}$,
  $\chi\rightarrow\tau^+\tau^-$,
  $\chi\rightarrow \mu^{+}\mu^{-}$, and $\chi\rightarrow \nu\overline{\nu}$, respectively.}
  \label{fig:8}
\end{figure}

\begin{figure}
  \centering
  \includegraphics[width=0.85\textwidth]{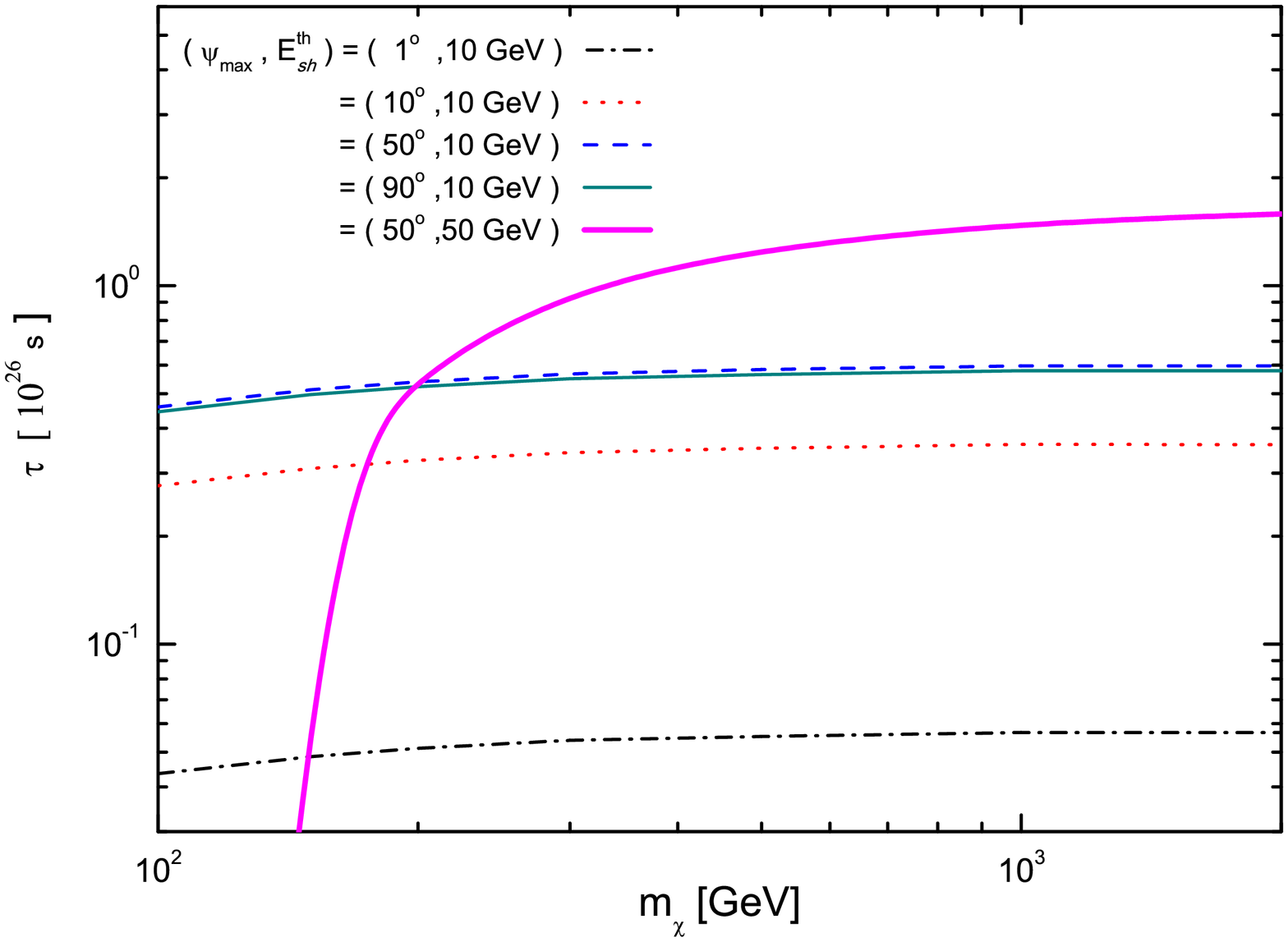}
  \vskip 0.5cm
  \caption{ The required DM decay time $(\chi\rightarrow\mu^{+}\mu^{-})$
  as a function of $m_{\chi}$ such that the neutrino signature from DM decays can be
  detected at the $2\sigma$ significance in five years for cascade events.
  Results corresponding to different $\psi_{\rm max}$ are
  presented. For comparison, we also show the result with $E_{sh}^{\rm th}=50$ GeV
  and $\psi_{\textrm{max}}=50^{\circ}$ \cite{Erkoca:2010}.}
  \label{fig:9}
\end{figure}

Dwarf spheroidal satellite galaxies of the Milky Way are DM-dominated systems which do not
have active star formation or detected gas content~\cite{Mateo:1998,Grcevich:2009}.
Satellite galaxies are among the best targets to search for DM signals in gamma rays
because of small background from astrophysical sources and a favorable signal
to noise ratio. In Ref.~\cite{Ackermann:2011}, Fermi-LAT collaboration derive
upper limits on DM annihilation cross sections by applying a joint likelihood analysis to 24 months of data from
10 satellite galaxies with uncertainties on the dark matter distributions in the satellite galaxies taken into account.
We present  the dSphs
constraint on $\chi\chi\to \mu^+\mu^-$ annihilation cross section at $95\%$ C.L. in Fig.~6.
For $m_{\chi}>300~\textrm{GeV}$, the expected DeepCore sensitivity curves are below that set by 
dSphs constraint for both $\psi_{\textrm{max}}=50^{\circ}$  and $\psi_{\textrm{max}}=10^{\circ}$. For $m_{\chi}< 300$ GeV, the dSphs constraint is comparable to the expected DeepCore sensitivity
with $\psi_{\textrm{max}}=10^{\circ}$. Thus it is stronger than  the expected DeepCore sensitivity
with $\psi_{\textrm{max}}=50^{\circ}$ in this DM mass range.

\begin{figure}
  \centering
  \includegraphics[width=0.85\textwidth]{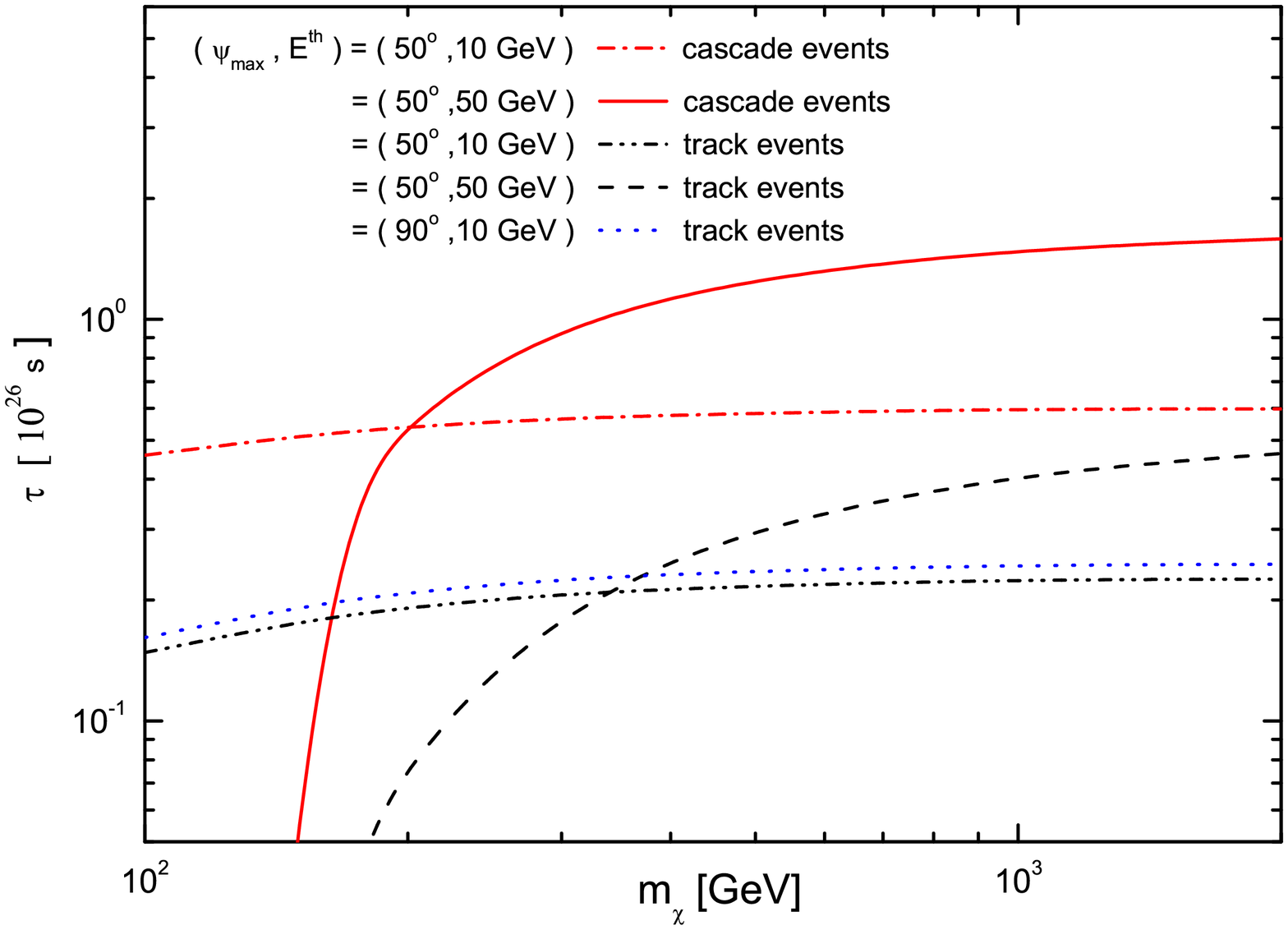}
  \vskip 0.5cm
  \caption{ The required DM decay time
  $(\chi\rightarrow\mu^{+}\mu^{-})$ as a function of $m_{\chi}$
  such that the neutrino signature from DM decays can be detected at the $2\sigma$
  significance in five years for track and cascade events.}
  \label{fig:10}
\end{figure}

We also present Fermi-LAT's dSphs constraint and galactic gamma ray  constraint on $\chi\chi\to b\bar{b}$ annihilation cross section in Fig.~7.
By comparing Figs.~6 and 7, we note that Fermi-LAT data gives more stringent constraint on $\chi\chi\to b\bar{b}$ mode  than its constraint on $\chi\chi\to \mu^+\mu^-$ mode.   
This is in contrast to the DeepCore case as one can see from Fig.~2.  While the neutrino spectrum through $\chi\chi\to \mu^+\mu^-$ is harder than that through $\chi\chi\to b\bar{b}$,
the gamma ray spectra through the above annihilations behave differently. The gamma ray spectrum from $\chi\chi\to b\bar{b}$ dominates over that from $\chi\chi\to \mu^+\mu^-$  
for most of the range of $x=E_{\gamma}/m_{\chi}$~\cite{Cirelli:2010xx}. 
The expected $2\sigma$ sensitivity on $\chi\chi\to b\bar{b}$ channel by the DeepCore detector
with $\psi_{\textrm{max}}=10^{\circ}$ is also shown in Fig.~7. One can see that the expected DeepCore sensitivity
on this channel is weaker than all existing constraints presented here. 

\begin{figure}
  \centering
  \includegraphics[width=0.85\textwidth]{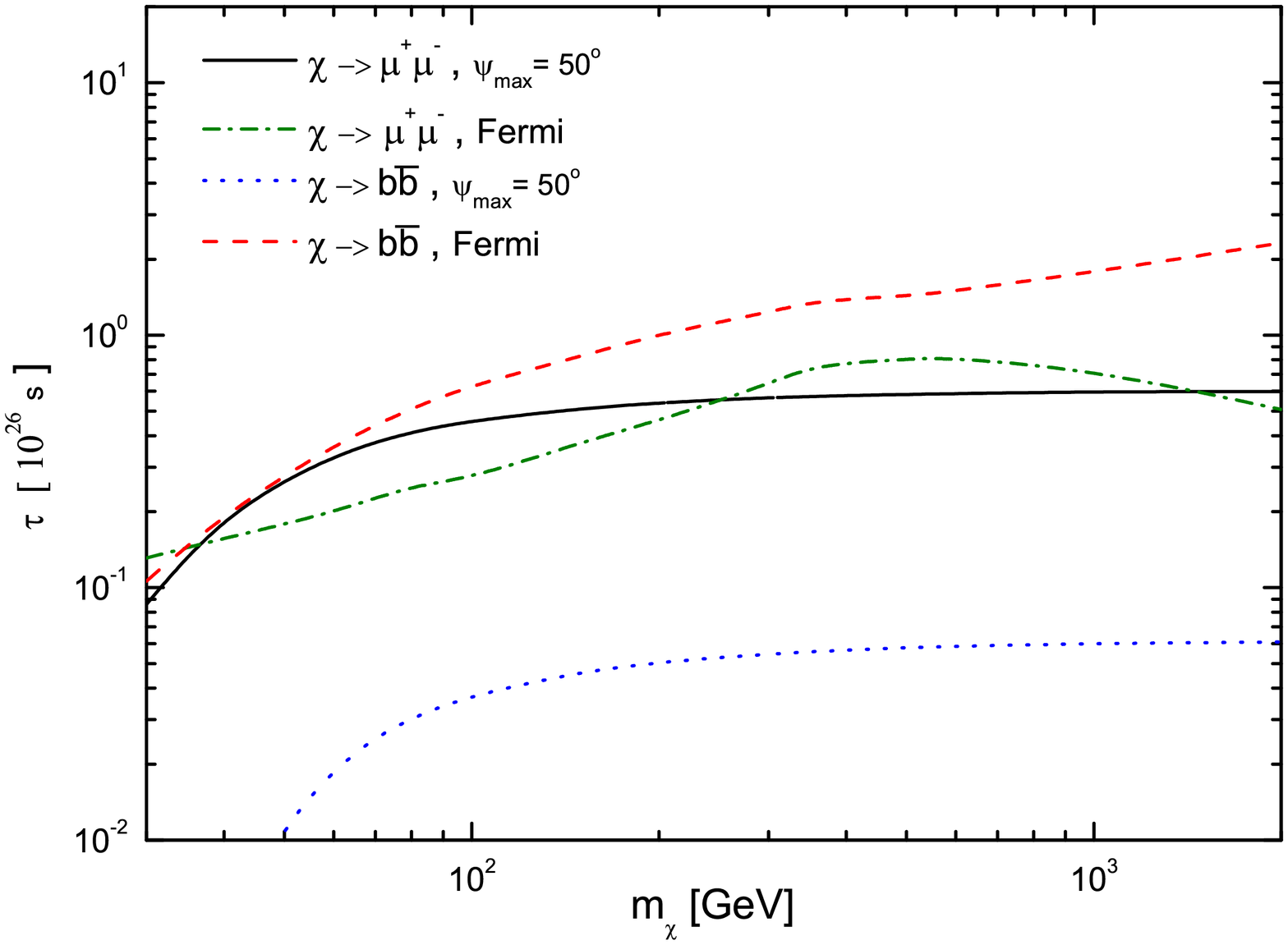}
  \vskip 0.5cm
  \caption{
  The dot-dashed line and dashed line are the decay time constraints for $\chi\to \mu^+\mu^-$ and $\chi\to b\bar{b}$
  channels due to Fermi observations for the region $|b| > 10^{\circ}$ plus a $20^{\circ}\times20^{\circ}$ at GC,
  assuming the NFW profile ~\cite{Fermi:2012}. The solid line and dotted line are our expected $2\sigma$ sensitivities
  for $\chi\to \mu^+\mu^-$ and $\chi\to b\bar{b}$ channels with
  $E_{sh}^{\textrm{th}}=10~\textrm{GeV}$ and $\psi_{\textrm{max}}=50^{\circ}$.  }
  \label{fig:11}
\end{figure}

In addition to studying DeepCore sensitivities on DM annihilation channels,
we also study sensitivities on DM decay time for $\chi\to b\bar{b}, \ \tau^+ \tau^-$,
$\mu^+\mu^-$ and $\nu\overline{\nu}$ channels. Figure~8 shows the required DM decay time for reaching  $2\sigma$ detection significance in five years on cascade events from each channel with
the threshold energy $E_{sh}^{\textrm{th}}=10~\textrm{GeV}$ and a cone half-angle $\psi_{\textrm{max}}=50^{\circ}$. Non-detection of such a signature would
then exclude the parameter region below the curve at the $2\sigma$ level.  We shall see later 
that the DeepCore sensitivity to DM decay time is not improved by considering $\psi_{\textrm{max}}$ smaller than $50^{\circ}$. Comparing various DM decay modes, one can see that
$\chi\rightarrow \nu\overline{\nu}$ channel requires the lowest decay rate (longest decay time) to avoid a
$2\sigma$ detection significance in five years of DeepCore data taking. We note that the energy-dependent effective volume  $V_{\rm{casc}}(E)$~\cite{Collaboration:2011ym}
downgrades the sensitivity to $\chi\to \nu\bar{\nu}$ decay time by roughly a factor of 1.5 at $m_{\chi}=30$ GeV, which corresponds to $E_{\nu}= 15$ GeV. However, the sensitivity to
 $\chi\to \nu\bar{\nu}$ obtained with $V_{\rm{casc}}(E)$
is better than that presented in Fig.~8 for $m_{\chi}> 80$ GeV, which corresponds to $E_{\nu}> 40$ GeV.  For $\chi\to \mu^+\mu^-$ mode,  the sensitivity at $m_{\chi}=30$ GeV is lower by a factor of $\sqrt{3}\approx 1.7$ by applying $V_{\rm{casc}}(E)$, since such a $m_{\chi}$ corresponds to lowest possible neutrino energy $E_{\nu}=10$ GeV. However, $V_{\rm{casc}}(E)$ shall enhance the sensitivity for $m_{\chi}> 240$ GeV, which corresponds to $E_{\nu}=40$ GeV, assuming $E_{\nu}\simeq m_{\chi}/6$. 
The correction factor for $\chi\to \tau^+\tau^-$ is similar to that for $\chi\to \mu^+\mu^-$, while the hadronic mode $\chi\to b\bar{b}$ requires different correction factor for the same 
$m_{\chi}$. Once more, we do not address such a correction as the DeepCore detector is relatively insensitive to $\chi\to b\bar{b}$.        

Next, we show how the DeepCore sensitivity on DM decay time varies with
the chosen cone half-angle and threshold energy. We use
$\chi\to \mu^+\mu^-$ channel to illustrate these effects.
In Fig.~9, we present the required DM decay time
$(\chi\rightarrow\mu^{+}\mu^{-})$ as a function of $m_{\chi}$ for reaching  $2\sigma$ detection significance in five years
for different cone half-angle $\psi_{\textrm{max}}$. The sensitivity curve rises as
$\psi_{\textrm{max}}$ increases from $1^{\circ}$ to $50^{\circ}$. As
$\psi_{\textrm{max}}$ increases in this cone half-angle range, the DM event rate  
increases faster than that of atmospheric
background. However, the sensitivity is not further improved by increasing $\psi_{\textrm{max}}$ from $50^{\circ}$ to $90^{\circ}$.
We also show the required DM decay time for a $2\sigma$ detection significance in five years
with $E_{sh}^{\textrm{th}} = 50~\textrm{GeV}$ and $\psi_{\textrm{max}} = 50^{\circ}$
for comparison. It has been pointed out in Ref.~\cite{Erkoca:2010} that
$\psi_{\textrm{max}} = 50^{\circ}$ gives rise to the highest sensitivity on
DM decay time for $E_{sh}^{\textrm{th}}= 50~\textrm{GeV}$.
We note that the sensitivity on DM decay time is improved by lowering
$E_{sh}^{\textrm{th}}$ from $50$ GeV to $10$ GeV for $m_{\chi}< 200$ GeV.

It is of interest to compare sensitivities on  DM decay time given by
cascade events and track events. In Fig.~10, one can see that
cascade events provide better sensitivity on DM decay time than that given
by track events for the same threshold energy $E^{\textrm{th}}$ and $\psi_{\textrm{max}}$.
In the same figure we also show the sensitivity given by track events for $\psi_{\rm{max}}=90^{\circ}$. Such a $\psi_{\rm{max}}$ renders 
the best sensitivity  for track events. However this sensitivity is
still poorer than those given by cascade events.

In Fig.~11, we present decay time lower limits for $\chi\to \mu^+\mu^-$ and $\chi\to b\bar{b}$
channels obtained from the diffuse gamma-ray spectrum for the region $|b| > 10^{\circ}$ plus a $20^{\circ}\times20^{\circ}$ square region centered 
at GC, assuming the NFW profile. They are taken from Ref.~\cite{Fermi:2012},
with a  rescaling factor $3/4$ applied since we have adopted  a local density of $\rho_{\odot}=0.3~\textrm{GeV}/\textrm{cm}^{3}$ while Fermi-LAT analysis uses  $\rho_{\odot}=0.4~\textrm{GeV}/\textrm{cm}^{3}$.
In the same figure, we also show the expected $2\sigma$ sensitivities to $\chi\to \mu^+\mu^-$ and $\chi\to b\bar{b}$
decay time by the DeepCore detector with $\psi_{\textrm{max}}=50^{\circ}$ for comparisons.
For the DeepCore detector, the sensitivity to the $\chi\to \mu^+\mu^-$ decay time is better than that
to the $\chi\to b\bar{b}$ decay time, since the neutrino spectrum in the former channel is harder than the one in the latter channel. On the other hand, the Fermi-LAT data in general gives more stringent constraint
on $\chi\to b\bar{b}$ decay time than its constraint on the decay time of $\chi\to \mu^+\mu^-$.  This is because that the gamma ray spectrum from the former channel dominates over  the one from the latter channel for most of  the range of $x=E_{\gamma}/m_{\chi}$, as the DM annihilation case.  
If DM decays predominantly  into $\mu^+\mu^-$, one can see  that 
the decay time sensitivity expected at DeepCore is comparable to the constraint given by Fermi-LAT data. This conclusion is not altered by adopting the energy-dependent effective volume
$V_{\rm{casc}}(E)$~\cite{Collaboration:2011ym}. 
On the other hand, if DM decays predominantly to $b\bar{b}$,
the expected DeepCore sensitivity is much poorer than the constraint from Fermi-LAT data. 

\section{Discussions and Conclusions}
In this paper, we have evaluated sensitivities of IceCube DeepCore detector to neutrino cascade events induced by DM annihilations and decays in galactic halo. We focus on the scenario of small DM mass and the threshold energy for the cascade events is taken to be $10$ GeV. The event rate of background atmospheric neutrinos 
is calculated with $\nu_{\mu}\to \nu_{\tau}$ oscillations taken into account for neutrino energies less than $40$ GeV. The signal event rate is calculated by taking NFW profile 
for DM density distribution in the galactic halo. 
Among all DM annihilation and decay channels, the annihilation mode $\chi\chi\to \nu\bar{\nu}$ and the decay mode $\chi\to \nu\bar{\nu}$ provide the best search sensitivity, while 
the search sensitivity provided by the annihilation mode $\chi\chi\to b\bar{b}$ and the decay mode $\chi\to  b\bar{b}$ is the poorest. 

It is important to compare the expected sensitivities of  DeepCore detector to DM annihilation cross section and decay time with the existing constraints on the same quantities
obtained by cosmology and Fermi-LAT gamma-ray observations. It is seen that the Fermi-LAT constraints on  $\chi\chi\to b\bar{b}$ and $\chi\to  b\bar{b}$ are much stronger 
than the expected DeepCore sensitivities to the same channels. Hence, if DM predominantly annihilates or decays into $b\bar{b}$, the DeepCore detector is not expected to observe 
neutrino signature induced by DM in the galactic halo. 
For leptonic final states, one can see from Fig. 6 that the expected DeepCore sensitivity to $\chi\chi\to \mu^+\mu^-$ annihilation cross section is stronger than 
the Fermi-LAT constraint on the same channel  based upon gamma-ray data from the galactic halo. However, the former sensitivity with the current angular resolution of cascade events is 
slightly weaker than both the dSphs constraint and the constraint obtained from WMAP and ACT results on CMB anisotropy. On the other hand, the DeepCore sensitivity can be improved with an improved 
angular resolution for cascade events.
In Fig. 11, we also see that the expected DeepCore sensitivity to $\chi\to \mu^+\mu^-$ decay time is comparable to the Fermi-LAT constraint  based upon gamma-ray data from the galactic halo.
From the above comparisons, there remain slight possibilities to observe DM induced neutrino signature from the galactic halo provided DM annihilates or decays predominantly into leptons.  

It should be noted that Fermi-LAT and CMB data does not directly set limits on $\chi\chi\to \nu\bar{\nu}$ or $\chi\to \nu\bar{\nu}$ modes with monochromatic neutrinos. There exist models
\cite{Allahverdi:2009se,Falkowski:2009yz} in which 
$\chi\chi\to \nu\bar{\nu}$  and $\chi\to \nu\bar{\nu}$ are dominant annihilation and decay modes, respectively. In the annihilation case \cite{Allahverdi:2009se} for example, the DM candidate can be the lightest right-handed (RH) sneutrino in a $U(1)_{B-L}$ extension of  the minimal supersymmetric standard model. RH sneutrinos 
annihilate into a pair of RH neutrinos. Each of these RH neutrinos then decays into ordinary left-handed (LH) neutrino and a neutral Higgs boson while the decay of RH neutrino into charged final states $l^{\pm}h^{\mp}$ is typically forbidden in such a model. In the case that the mass difference between RH sneutrinos and RH neutrinos is small, RH neutrinos are produced non-relativistically 
by DM annihilations. Hence  LH neutrinos  produced by the decays of RH neutrinos are approximately monochromatic with an energy around half of the DM mass. Therefore the sensitivity of  IceCube DeepCore to this type of models can be read off from the $\chi\chi\to \nu\bar{\nu}$ curve in Fig. 2 with the shift $(\langle\sigma\upsilon\rangle,m_{\chi})\to (\langle\sigma\upsilon\rangle,2m_{\chi})$.

Before closing, we comment on the detection of neutrino signature induced by DM in the galactic halo with neutrino telescopes in the northern hemisphere. Generally a neutrino telescope in such a location has advantages in detecting track-like neutrino events. First,     
the telescope's effective volume for upward going track events is enhanced by the muon range. This effect is particularly significant for energetic track events originated from high-energy muon neutrinos. 
Second, a neutrino telescope in the northern hemisphere naturally suppresses atmospheric muon background while IceCube needs to use optical modules located at the outer region for vetoing the same background~\cite{Abbasi:2012ws}. Following these arguments, it is interesting to see if a detector array in the northern hemisphere with the size of DeepCore detector has a significantly better sensitivity  than the current DeepCore detector surrounded by IceCube strings.  Since the latter also has a good veto capability, the former can gain only in the effective volume expected to be enhanced by the muon range. We note that the DeepCore array aims at detecting  neutrinos in the energy range $10\le E_{\nu}/{\rm GeV}\le100$. Muons induced by muon neutrinos in this energy range only travel around $50$ m for $E_{\mu}=10$ GeV and $400$ m for $E_{\mu}=100$ GeV. 
This does not significantly enhance the detector's effective volume in most cases since the height of the DeepCore detector is already around $350$ m. Therefore, given the existence of IceCube detector augmented by DeepCore in the South Pole, it is clear that neutrino telescopes in the northern hemisphere only have advantages in detecting neutrino signature from heavier DM. In this regard, the acceptance for track events in KM3NeT as a function of neutrino energy has been estimated \cite{KM3NeT}. With this acceptance, we also calculate the sensitivity of KM3NeT to DM annihilation cross section  $\langle\sigma(\chi\chi\to \mu^+\mu^-)\upsilon\rangle$ in the galactic halo. 
For $\psi_{\rm max}=10^{\circ}$ and $m_{\chi}=200$ GeV, the KM3NeT $2\sigma$ sensitivity to $\langle\sigma(\chi\chi\to \mu^+\mu^-)\upsilon\rangle$ in 5 years  is comparable to the 
CMB constraint on this channel~\cite{Galli:2011}. 
However, for $m_{\chi}=1$ TeV, CMB constraint gives  $\langle\sigma(\chi\chi\to \mu^+\mu^-)\upsilon\rangle$ no greater than $10^{-23}$
cm$^3$s$^{-1}$ whereas the KM3NeT $2\sigma$ sensitivity on the same annihilation channel can reach to  $1.5\times 10^{-24}$cm$^3$s$^{-1}$ in 5 years. 

In conclusion, we have made detailed comparisons between IceCube DeepCore sensitivities and other existing constraints on various DM annihilation and decay channels. The prospect for DeepCore detector to observe  neutrino signature induced by DM in the galactic halo has been discussed.  We have also mentioned the expected performance of KM3NeT on this observation.     

\section*{Acknowledgments}
We thank Y.-H. Lin for useful discussions and C. Rott for bringing Ref.~\cite{Collaboration:2011ym} to our attention. F.F.L. is supported by the grant from Research and Development Office, National Chiao-Tung University, G.L.L. is supported by National Science Council
of Taiwan under Grant No. 99-2112-M-009-005-MY3,
Y.S.T. is funded in part by the Welcome Program of the Foundation for Polish Science.

\end{document}